# Pulse-Mode Operation and Reliability of BEOL-Compatible Ferroelectric Non-Volatile Capacitive Memories with Amorphous Oxide Semiconductor Channels

Junmo Lee, *Member, IEEE*, Chengyang Zhang, *Member, IEEE*, Tae-Hyeon Kim, *Member, IEEE*, Suman Datta, *Fellow, IEEE*, Shimeng Yu, *Fellow, IEEE*

***Abstract*— Non-volatile capacitive memories (nvCAPs) exhibiting AC "small-signal" capacitance on/off ratio ($C_{on}/C_{off}$) with non-destructive read have emerged as a promising device for next-generation memory paradigms. Recently, BEOL-compatible ferroelectric nvCAPs with an amorphous oxide semiconductor channel have been reported, suggesting the possibility of monolithic 3D integration of nvCAPs on top of CMOS. So far, the characterization studies on oxide-channel ferroelectric nvCAPs have been done using dual DC sweep C-V measurements which are typically performed over a time scale of a few seconds. However, non-volatile memory arrays typically require nvCAPs to operate under pulse-mode. It is thus crucial to advance understanding of the behavior of oxide-channel ferroelectric nvCAPs under pulse-mode operation, governed by the unique interplay between ferroelectric layer and oxide channel physics. In this study, we provide a systematic study of the pulse-mode operation of ferroelectric nvCAPs with an amorphous oxide semiconductor channel, including its pulse-based write characteristics and reliability characteristics. We examine overlap area, wake-up and pulse-width dependent $C_{on}$ and $C_{off}$ writing characteristics under pulse-mode. Further, we suggest the importance of optimizing ferroelectric depolarization for $C_{on}$ retention, while reducing read-after-delay for $C_{off}$ retention under pulse-mode. Lastly, non-destructive read operation for $>10^9$ read stress cycles at $|V_{read}|$=1V is demonstrated.**

## I. INTRODUCTION

The increasing computational demand of artificial intelligence (AI) has spurred extensive research into developing emerging memory devices for novel computing paradigms. Numerous types of memories, such as resistive random access memory (RRAM), phase change memory (PCM), and ferroelectric field-effect transistor (FeFET) have been actively investigated for applications in current domain compute-in-memory (CIM) [1]. More recently, non-volatile capacitive memory (nvCAP) has emerged as a promising memory candidate for next-generation charge-domain CIM [2-5]. The foundational principle of nvCAP is its AC small signal capacitance memory window at 0V. Such a mechanism allows non-destructive read-out of the two capacitance states ($C_{on}$, $C_{off}$) with a small read voltage (<200 mV), allowing low IR drop, significantly lower static power consumption, and read disturbance mitigation compared against existing non-volatile memories. To further enhance its area efficiency through monolithic 3D integration, BEOL-compatible ferroelectric nvCAPs with amorphous oxide semiconductor channels, such as W-doped $In_2O_3$ (IWO) and IGZO have been proposed [6, 7]. One of the main challenges in developing BEOL-compatible oxide-channel ferroelectric nvCAPs is achieving sufficient capacitance on/off ratio ($C_{on}/C_{off}$) while maintaining BEOL-compatible process temperature (<450 °C) throughout the entire fabrication process [6]. Prior work on ferroelectric nvCAPs with an ALD IWO channel [6] reported comparable $C_{on}/C_{off}$ to that of silicon-channel ferroelectric nvCAPs, by means of W-doping control and overlap area engineering. However, the results were obtained from dual DC sweep C-V measurement with a slow

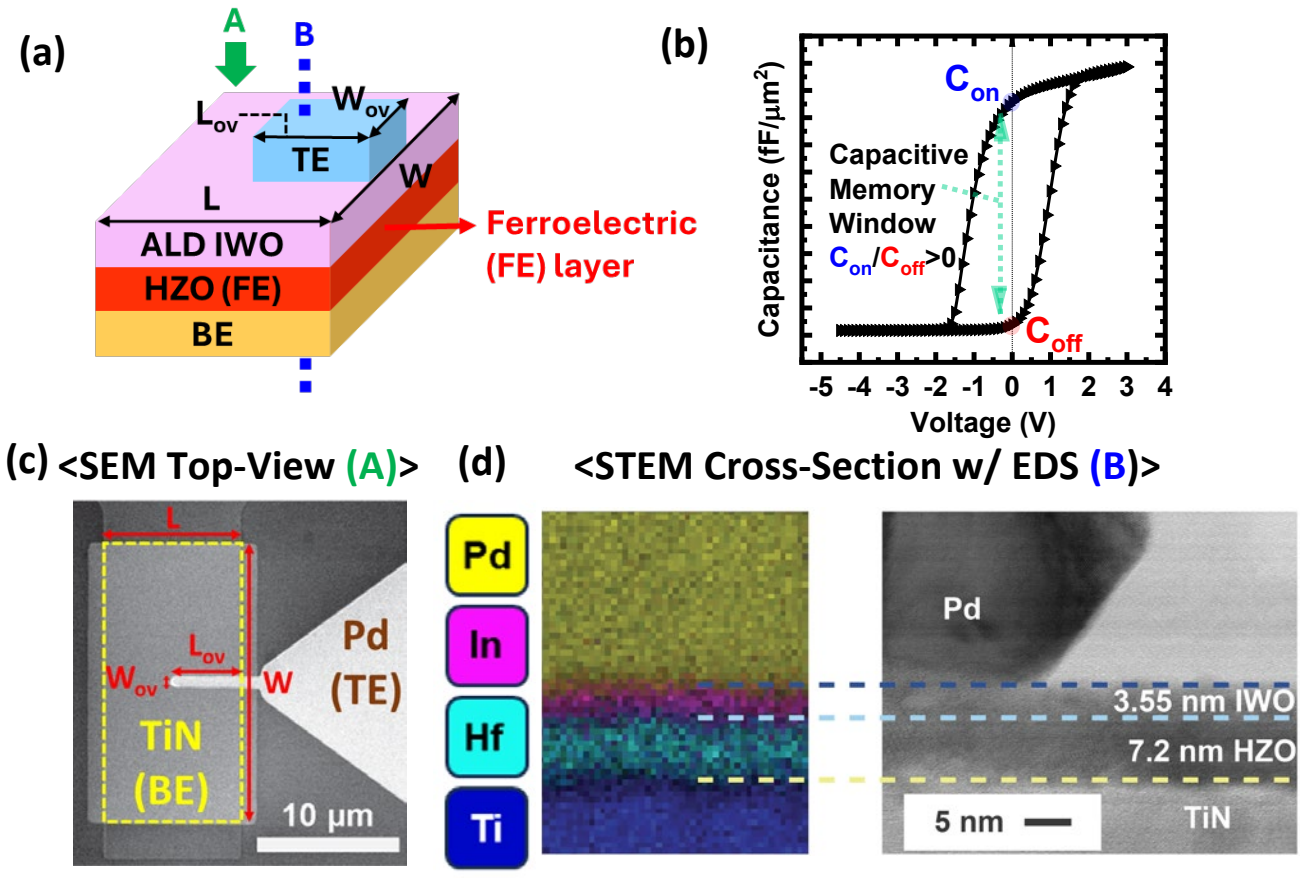

Fig. 1. (a) Schematic illustration of the BEOL-compatible ALD IWO nvCAP studied in this work and (b) its typical C-V characteristics. (c) Top view SEM image and (d) cross-sectional STEM image with EDS elemental mapping of the ALD IWO nvCAP.

This work was supported by PRISM and CHIMES, two of the semiconductor research corporation (SRC) centers. The authors are with the School of Electrical and Computer Engineering, Georgia Institute of Technology, Atlanta, GA 30332 USA (e-mail: junmolee@gatech.edu, shimeng.yu@ece.gatech.edu)

DC sweep rate (~6.7 mV/ms), which does not truly capture the behavior of oxide-channel ferroelectric nvCAPs under pulse-mode operation in a practical memory array [8, 9].

This work systemically examines the pulse-mode operation of oxide channel ferroelectric nvCAPs, focusing on their pulse-mode writing, retention, write endurance, and read endurance characteristics under different device geometries and operating conditions. We introduce a BEOL-compatible ferroelectric nvCAP with an ALD IWO channel fabricated with an improved process flow from our prior work [6]. We outline the governing device physics arising from unique interplay between the ferroelectric (FE) and channel layer, leading to several non-idealities under pulse-mode. The device operation and polarization dynamics under pulse-mode are carefully examined with respect to the underlying physics, aiming to provide useful device-level insights for further oxide-channel ferroelectric nvCAP improvements.

## II. Ideal and Non-Ideal Operational Principles of BEOL-Compatible ALD IWO nvCAP

Fig. 1 (a)-(b) shows a schematic illustration and typical C-V characteristics of a BEOL-compatible ferroelectric nvCAP with an ALD IWO channel (ALD IWO nvCAP) studied in this work. Fig. 1 (c) and (d) show the top view SEM and cross-sectional STEM images with EDS elemental mapping of the ALD IWO nvCAP, respectively.

Ferroelectric nvCAPs with an oxide channel operating under "pulse-mode" exhibit several *non-idealities* leading to deviations of measured $C_{on}$ and $C_{off}$ from their ideal values ($C_{on,ideal}$ and $C_{off,ideal}$) obtained under ideal conditions, typically assumed in the literature. It is thus crucial to examine the physical mechanisms of program (PGM) and erase (ERS) states of the oxide-channel ferroelectric nvCAPs under its ideal (Fig. 2(a)) and non-ideal operations (Fig. 2(b)-(c)). It should be noted that all the capacitance expressions (e.g., $C_{on}$, $C_{off}$) used throughout this paper are in normalized units (fF/$\mu m^2$).

When a program voltage ($V_{PGM}$) is applied to the nvCAP, the channel becomes accumulated above up-polarized FE domains. Ideally, assuming the accumulated channel behaves as a "metal" with zero resistance, the channel becomes accumulated across the entire device area, resulting in $C_{on,ideal}$ of $C_{HZO}$ (Fig. 2(a)). In practice, however, non-zero resistance of the oxide channel creates RC delay along the lateral direction of the non-overlap region. Thus, the measured $C_{on}$ can be expressed as $(A_{acc,PGM}/A_{BE})\times C_{HZO}$ and can vary depending on the PGM pulse conditions, overlap area conditions and the channel properties (Fig. 2(b)). Furthermore, the ferroelectric depolarization causes the measured $C_{on}$ to vary over time, which is not reflected in dual DC sweep C-V measurement results.

When an erase voltage ($V_{ERS}$) is applied to the nvCAP, negative voltage drop occurs in FE domains through holes supplied by TE and positively charged donor states at the HZO/IWO interface providing positive compensating charges [10]. Assuming full depletion ($C_{IWO}=C_{fulldep}$, where $C_{fulldep}$ is the capacitance of the IWO layer that is fully depleted) of the channel across the entire device area after ERS, which has been typically assumed in the literature, $C_{off}$ can be expressed as the

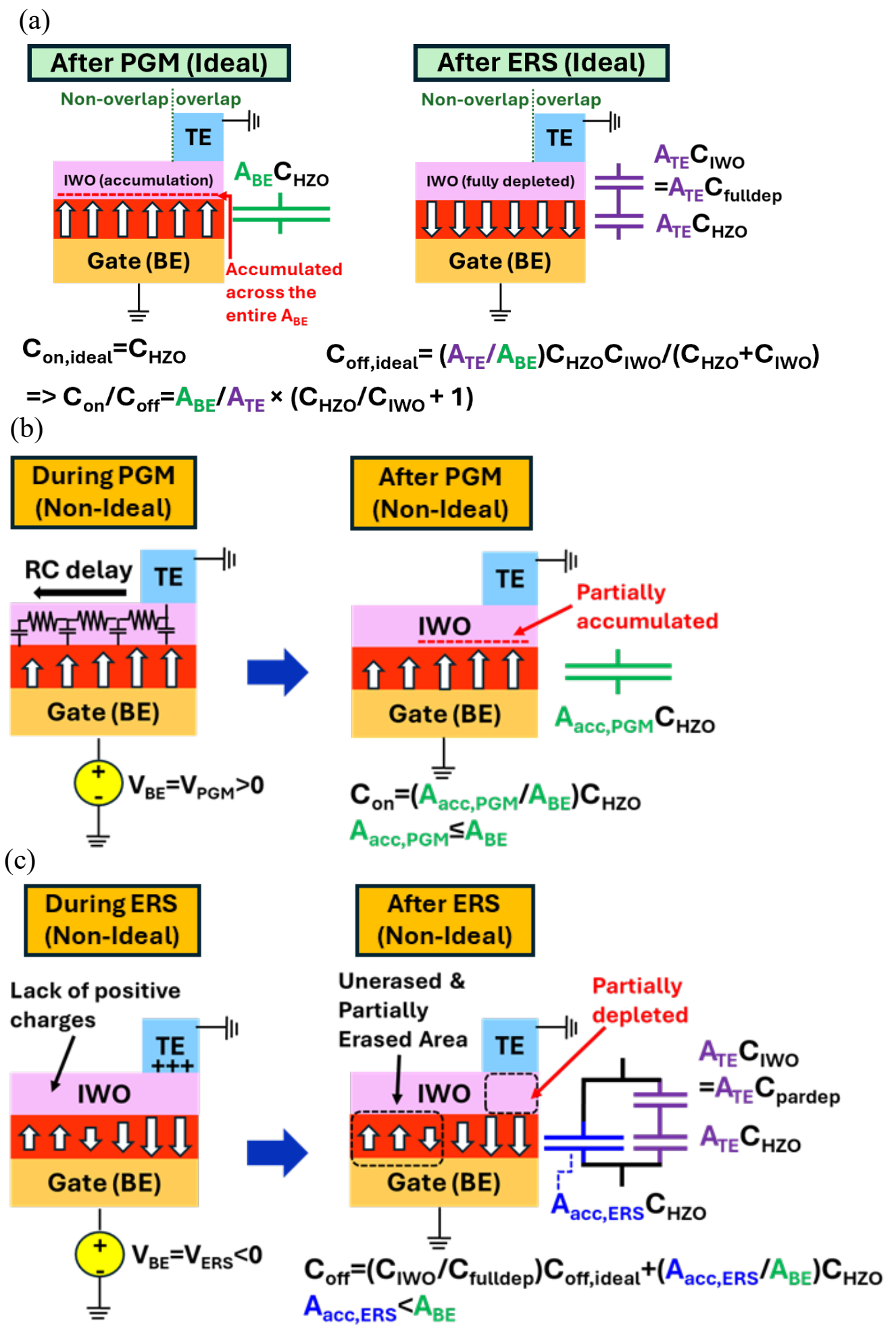

Fig. 2 (a) The derivation of $C_{on}$ and $C_{off}$ under ideal oxide-channel ferroelectric nvCAP operation assumptions. (b) The derivation of $C_{on}$ considering the non-idealities of oxide-channel ferroelectric nvCAPs. (c) The derivation of $C_{off}$ considering the non-idealities of oxide-channel ferroelectric nvCAPs.

series connection of $A_{TE}\times C_{IWO}$ and $A_{BE}\times C_{HZO}$ normalized by the device area ($A_{BE}$) (Fig. 2(a)).

However, there are several non-idealities of ERS originating from the inherent properties of oxide material, which becomes the key motivation of this study (Fig. 2(c)). In the case of ERS, the positive compensating charge for polarization is significantly smaller in non-overlap areas compared to overlap areas due to hole-deficient oxide channel layer and long generation time associated with positive charged donor states [11, 12]. This results in incomplete ERS due to a large delay in negative voltage drop across HZO in the non-overlap area. This can cause the channel to be partially accumulated after ERS ($A_{acc,ERS}>0$) when the ERS pulse width is not large enough. Likewise, domain pinning of up-polarized FE domains through repeated write cycling can be a possible cause of $A_{acc,ERS}>0$. Moreover, the channel in the overlap region can be partially depleted ($C_{IWO}=C_{pardep}>C_{fulldep}$, where $C_{pardep}$ is the capacitance of the IWO layer that is partially depleted) even after polarization switching, depending on the W doping concentration (intrinsic channel carrier concentration) of the channel and the presence of up-polarized FE domain pinning.

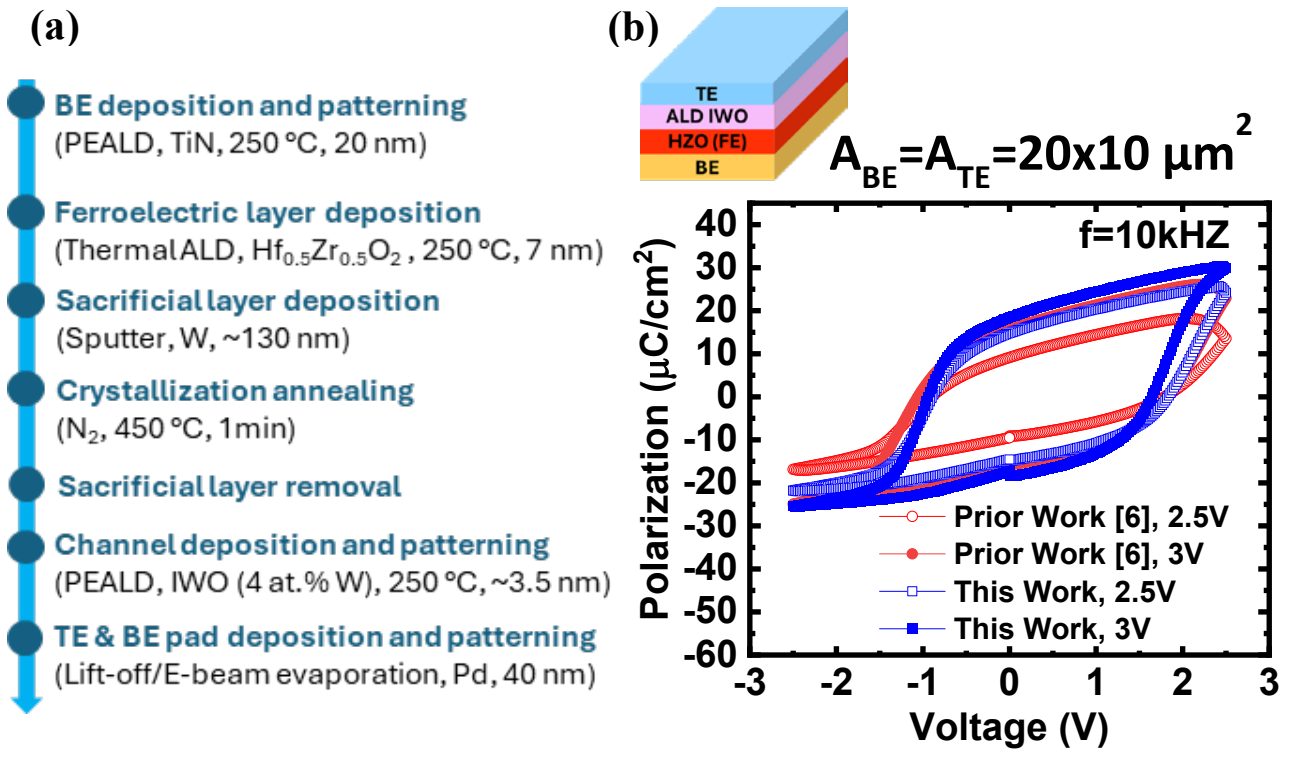

Fig. 3 (a). The process flow used for the fabrication of ALD IWO nvCAPs. (b) The obtained P-V characteristics of the ALD IWO nvCAP with symmetric TE and BE, with an area of 20x10 μm$^2$.

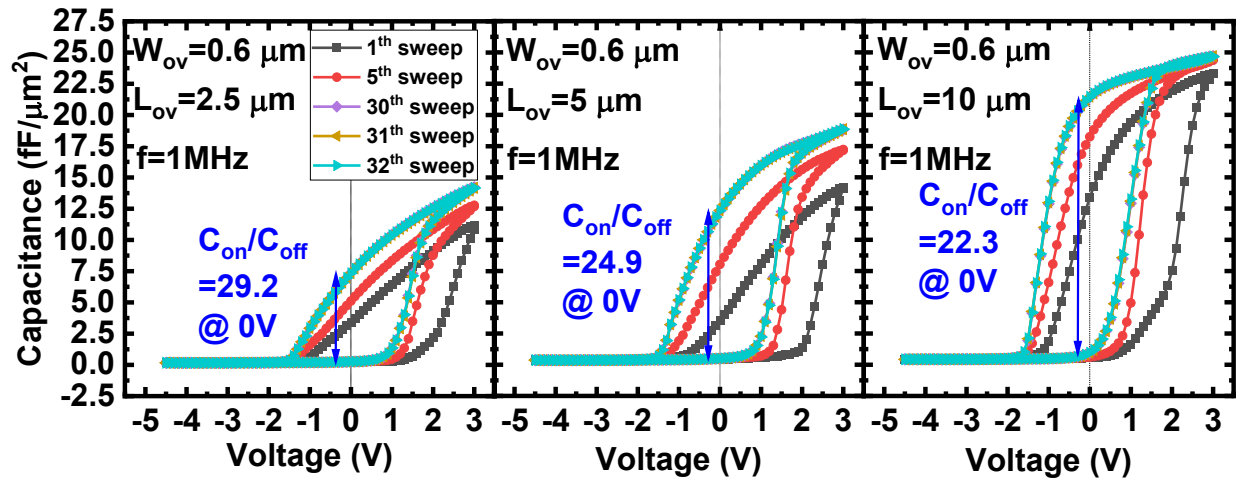

Fig. 4. C-V curves of ALD IWO nvCAPs obtained through bi-directional DC CV sweeps $W_{ov}/L_{ov}$ conditions are (a) 0.6/2.5 μm (b) 0.6/5 μm (c) 0.6/10 μm (W/L=20/10 μm).

Thus, the measured $C_{off}$ can widely vary depending on ERS pulse condition, interface/channel property, and the overlap conditions. More importantly, due to the ferroelectric polarization dynamics, $A_{acc,ERS}$ can vary over time, highlighting the importance of carefully evaluating the writing and reliability characteristics (retention, read-endurance and write-endurance) under pulse-mode for practical applications.

## III. Process Flow for BEOL-Compatible ALD IWO Ferroelectric nvCAPs

The ALD IWO nvCAP used in this study is fabricated following the process flow described in Fig. 3(a). W concentration is optimized to 4 at.% in the ALD super-cycling scheme established in our previous work [6]. In this study, we adopt an improved ferroelectric process over prior work [6] for enhanced ferroelectric performance (larger $P_r$). Sacrificial W layer thickness is increased from 80 nm to 130 nm to achieve better ferroelectric properties by applying larger mechanical stress to the HZO layer during the annealing process [13]. Fig. 3(b) shows the comparison of the P-V characteristics of an nvCAP with symmetric TE/BE between the prior work and this work. Here, the symmetric TE/BE design is compared to exclude the impact of the channel resistance on P-V characteristics. A higher $P_r$ is achieved in this work due to the improved process. In the following sections, we examine the pulse-mode operation of nvCAPs with asymmetric TE/BE designs, which exhibit higher $C_{on}/C_{off}$ than the symmetric TE/BE design.

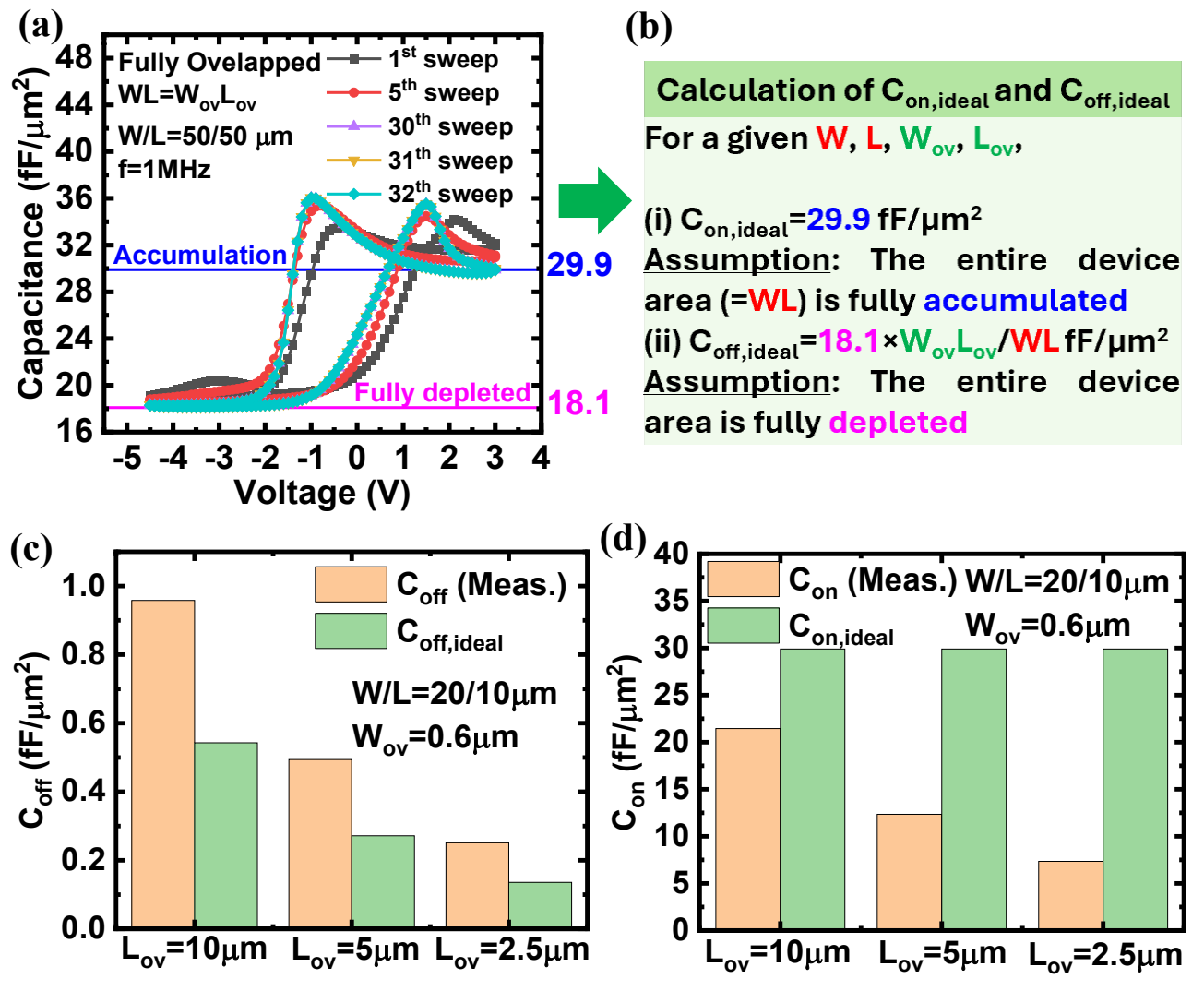

Fig. 5. (a) C-V curves of the ALD IWO nvCAPs with fully overlapped TE and BE at different sweep counts. (b) Calculation of $C_{on,ideal}$ and $C_{off,ideal}$ using the information from (a). (c) The comparison between DC sweep-based $C_{off}$ and $C_{off,ideal}$ (f=1 MHz). (d) The comparison between DC sweep-based $C_{on}$ and $C_{on,ideal}$ (f=1 MHz).

## IV. Impact of Write Pulse Conditions on $C_{ON}$ and $C_{OFF}$ of BEOL-Compatible ALD IWO nvCAPs

Prior works have examined writing characteristics of BEOL-compatible ferroelectric nvCAPs using dual DC sweep C-V measurement with a slow DC sweep rate [6, 7]. However, C-V measurement in typical measurement equipment is performed over a time scale of a few seconds (~2.2 sec with a DC sweep rate of ~6.7 mV/ms in our prior work [6]), which is comparable to applying ~1 s long write pulses to the nvCAPs for PGM/ERS under the pulse-mode operation. For practical applications requiring high clock frequency [14], it is essential to systematically examine pulse-based writing characteristics of oxide-channel ferroelectric nvCAPs and understand the difference from those obtained under bi-directional DC voltage sweeps.

### *A. $C_{on}$ and $C_{off}$ Characteristics Under Dual DC Sweep C-V Measurement*

Fig. 4 shows $C_{on}$ and $C_{off}$ of ALD IWO nvCAPs with different overlap parameters ($L_{ov}$) obtained through dual DC sweep C-V measurement at 25 °C. It should be noted that the total duration of the dual DC sweeps is set to be ~ 2.2 seconds (~1.1 sec to sweep from -4.5V to 3V, ~1.1 sec to sweep from 3V to -4.5V) with a DC sweep rate of ~6.7 mV/ms in our Keysight B1500A system. The gradual negative shift of the C-V curves over initial cycles ($1^{st}$~$30^{th}$ sweep) is an indication of increasing number of (i) woken up FE domains [15, 16], (ii) remaining up-polarized FE domains in the non-overlap region after weak ERS [11], (iii) positively charged defects at HZO/IWO interface causing negative $V_{th}$ shift, and (iv) up-polarized FE domain pinning [17, 18] over repeated DC sweeps. The C-V curve stabilizes after ~30 cycles, as (i)-(iv) saturate.

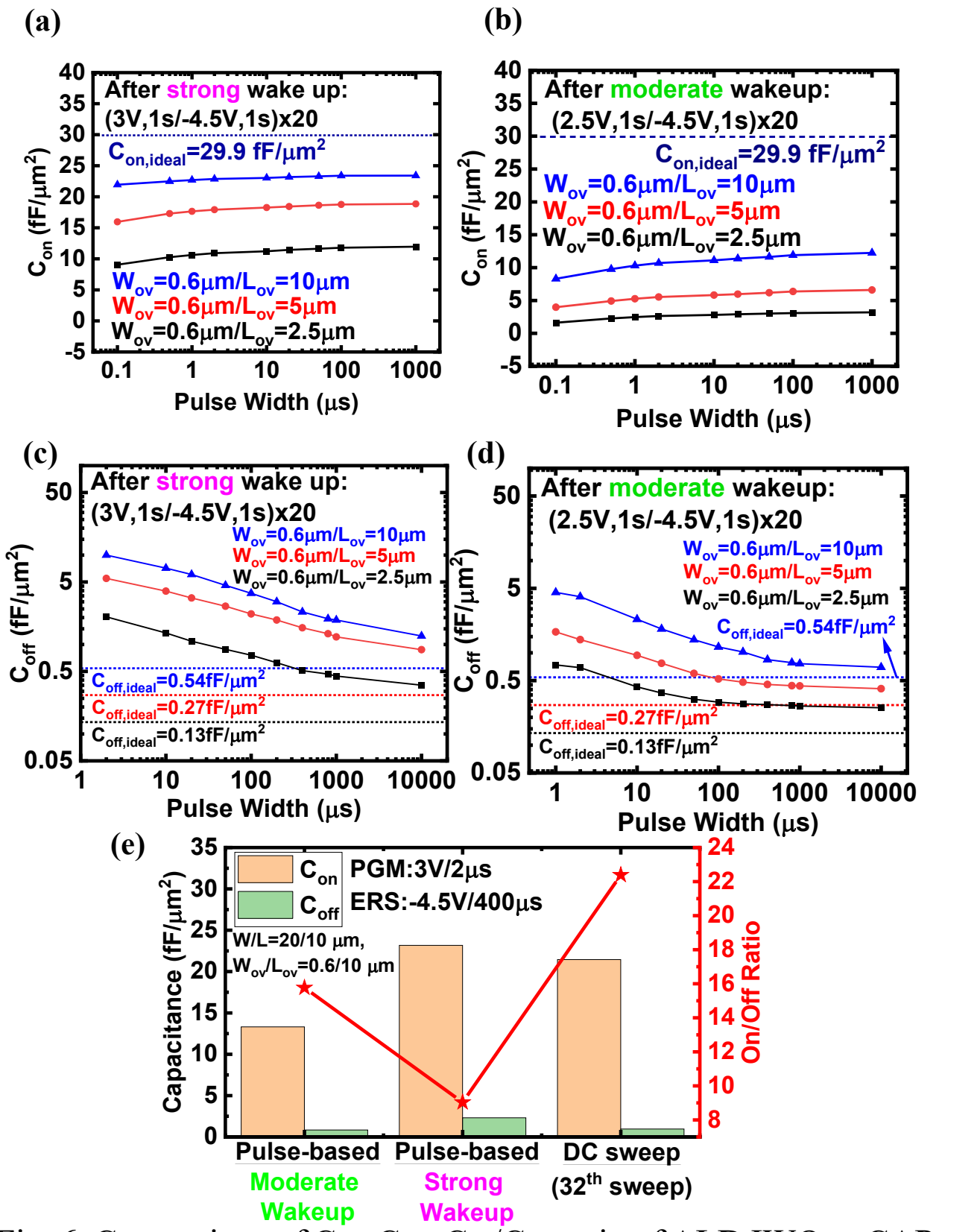

Fig. 6. Comparison of $C_{on}$, $C_{off}$, $C_{on}/C_{off}$ ratio of ALD IWO nvCAPs under different writing schemes and wake-up conditions. Plots of $C_{on}$ after (a) strong wake up pulses: 20 cycles of 3V,1s/-4.5V,1s, (b) moderate wake up pulses: 20 cycles of 2.5V,1s/-4.5V,1s. Plots of $C_{off}$ after (a) strong wake up pulses: 20 cycles of 3V,1s/-4.5V,1s, (b) moderate wake up pulses: 20 cycles of 2.5V,1s/-4.5V,1s. For all results in (a)-(e), W/L=20/10 μm and f=1 MHz.

We compare the stabilized $C_{on}$ and $C_{off}$ at the 32$^{nd}$ sweep with $C_{on,ideal}$ and $C_{off,ideal}$ calculated from the dual DC sweep C-V results of an ALD IWO nvCAP with symmetric (fully overlapped) TE and BE (Fig. 5(a)-(b)). It is shown that the lowest overlap area ($W_{ov}$=0.6 μm, $L_{ov}$=2.5 μm) leads to the highest $C_{on}/C_{off}$ ratio of ~29.2. Consistent with prior work [6], however, $C_{on}$ tends to degrade as the overlap dimension decreases from $L_{ov}$=10 μm to $L_{ov}$=2.5 μm. This is primarily attributed to the increasing RC delay (Fig. 2(b)) in the voltage drop across HZO layer in the non-overlap region during PGM, as the overlap area reduces. This leads to varying $A_{acc,PGM}$ with overlap area at the same PGM pulse width. The difference between $C_{on,ideal}$ and the measured $C_{on}$ thus increases as overlap area decreases. Meanwhile, the measured $C_{off}$ is found to be ~2× larger than the $C_{off,ideal}$ due to nonidealities such as partial depletion of the channel and incomplete ERS in the non-overlap area ($A_{acc,ERS}$>0) discussed in Fig. 2(c).

### *B. Pulse and Wake-up Condition Dependent $C_{on}$ and $C_{off}$ Characteristics*

We now examine $C_{on}$ and $C_{off}$ at 25 °C obtained through "pulse-based" writing under varying pulse widths as compared to those obtained through dual DC sweeps (Fig. 6). Considering the nature of oxide-channel nvCAPs exhibiting different C-V characteristics depending on the number of woken-up FE domains, two different wake-up pulse trains are applied to the BE of ALD IWO nvCAPs prior to $C_{on}$ and $C_{off}$ measurement (while TE is connected to ground): (i) moderate wake up: 20 cycles of (2.5V,1s)/(-4.5V,1s) (ii) strong wake up: 20 cycles of (3V,1s)/(-4.5V,1s). The pulse amplitudes for PGM and ERS are 3V and -4.5V, respectively. After wake-up, every PGM (ERS) pulse with a given pulse width is applied after a delay of <1 s following a reset pulse of −4.5 V/400 μs (3 V/20 μs) applied to the device. $C_{on}$ or $C_{off}$ is then measured using an AC small-signal amplitude of 0.1 V at 1 MHz, approximately 1 s after every PGM (ERS) pulse.

First, $C_{on}$ under pulse-based writing is examined. Figs. 6(a)-(b) show that the wake-up condition has a prominent influence on the $C_{on}$, while pulse width has comparably smaller effect. The stronger wake up conditions lead to a larger number of woken up FE domains in the non-overlap region, leading to more conductive channel formation during PGM, which in turn increases PGM efficiency. Increasing $C_{on}$ with increasing PGM pulse width observed for both (i) and (ii) can be explained by the RC delay along lateral direction of the non-overlap region. For both cases, comparable $C_{on}$ to that obtained from dual DC sweeps can be achieved with >1 μs PGM pulse width,

Next $C_{off}$ under pulse-based writing is examined in Figs. 6(c)-(d). The stronger wake-up conditions result in comparably larger number of incompletely erased or pinned up-polarized FE domains in the non-overlap region, resulting in larger $C_{off}$ and lower on-off ratio. It is noted that $C_{off}$ is much more sensitive to the pulse width conditions due to the inherently weak ERS caused by deficiency of positive charges in oxide materials. Generally, it can be observed that $C_{off}$ tends to rapidly increase below a pulse width of <400 μs. It should be noted that the pulse-based $C_{on}$ and $C_{off}$ presented so far —read ~1 s after each PGM/ERS pulse—do not capture their *time-dependent* characteristics. In the next section, we explain that $C_{off}$ will gradually decay to the $C_{off,ideal}$, while $C_{on}$ will gradually deviate from $C_{on,ideal}$ due to the ferroelectric depolarization.

Fig. 6 (e) shows a comparison of $C_{on}$ and $C_{off}$ of an ALD IWO nvCAP under pulse-mode operations ((i), (ii)) and dual DC sweeps. With stronger wake up pulses, $C_{on}$ from the pulse-mode is comparable to that from dual DC sweeps. However, $C_{on}/C_{off}$ is degraded when stronger wake up pulses are used, suggesting stronger wake-up is needed to target high $C_{on}$ while moderate wake-up condition is needed to improve $C_{on}/C_{off}$.

The results highlight the varying writing characteristics of oxide-channel ferroelectric nvCAPs depending on the pulse and wake up conditions. From an application point of view, they suggest the importance of co-optimization of overlap, pulse and wake-up condition to target desired $C_{on}$ and $C_{off}$.

## V. Reliability of BEOL-Compatible ALD IWO nvCAPs under Pulse-Mode Operation

Beyond the pulse-based write characteristics, an important question is the reliability of oxide-channel ferroelectric nvCAPs under the pulse-mode operation. Therefore, in this section, retention, write-endurance, and read-endurance characteristics of ALD IWO nvCAP under the pulse-mode operation are examined.

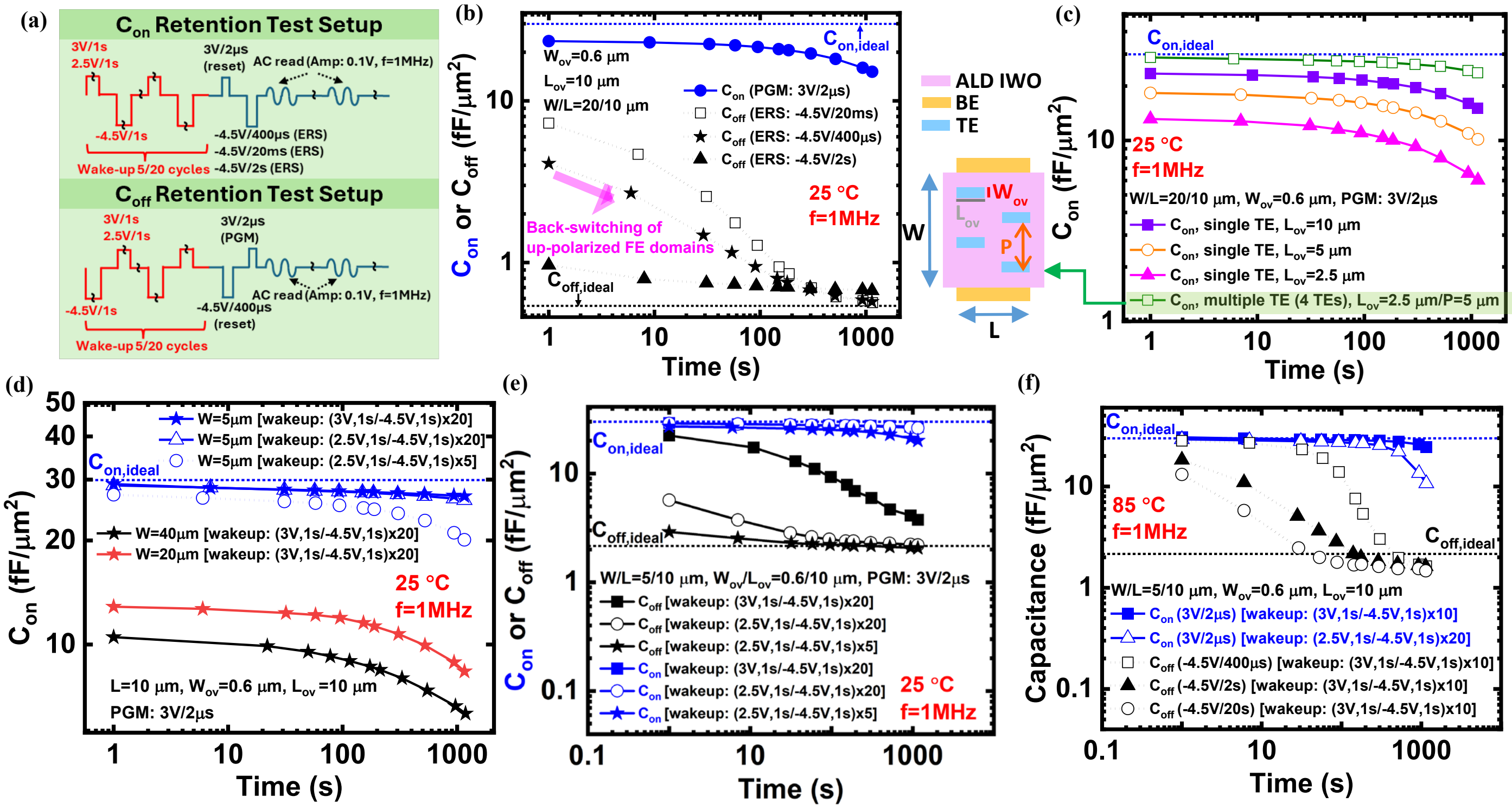

Fig. 7. Measured retention of $C_{on}$ and $C_{off}$. (a) The retention test setup and (b) retentions of $C_{on}$ and $C_{off}$ measured at different ERS pulse width conditions (25 °C). (c) Retention of $C_{on}$ at different overlap area conditions (25 °C). (d) Retention of $C_{on}$ at different device area conditions (25 °C). (e) Retention of $C_{on}$ and $C_{off}$ at different wake-up conditions (25 °C). (f) Retention of $C_{on}$ and $C_{off}$ at an elevated temperature (85 °C).

## *A. Retention*

Retention measurements are performed with a test setup described in Fig. 7(a). Fig. 7(b) shows the retention of $C_{on}$ and $C_{off}$ of an ALD IWO nvCAP with W/L=20/10 µm and $W_{ov}/L_{ov}$=0.6/10 µm. A noticeable trend is that both $C_{on}$ and $C_{off}$ decay over time, which is counterintuitive in terms of polarization dynamics suggesting decrease of $C_{on}$ and increase of $C_{off}$ over time. While the measured $C_{on}$ gradually deviates from $C_{on,ideal}$, $C_{off}$ gradually gets closer to $C_{off,ideal}$ over time.

This indicates only $C_{on}$ suffers from retention loss while $C_{off}$ is significantly affected by read-after-delay. The fundamental reason for such discrepancy between the retention trend of $C_{on}$ and $C_{off}$ is (i) hole-deficient channel layer leading to significantly weaker ERS compared to PGM and (ii) low enough carrier concentration of the channel leading to $C_{IWO}$=$C_{fulldep}$ above "non-polarized" FE domains. Upon applying an ERS pulse to a nvCAP reset to PGM state (t=1 s), measured $C_{off}$ is much higher than $C_{off,ideal}$ due to the incompletely erased up-polarized FE domains in the non-overlap region, causing fringing electric field between BE and TE. However, over time, the measured $C_{off}$ decays towards $C_{off,ideal}$ as up-polarized FE domains in non-overlap region depolarizes or back-switches to non-polarized domains (i.e., $A_{acc,ERS}$ decreases over time). This explanation is further evidenced by the fact that stronger ERS pulse conditions (larger pulse widths) lead to significantly reduced read-after-delay (Fig. 7(b)). The read-after-delay can be mitigated by using longer ERS pulse width (>20 ms). However, it should be noted that using such long pulses comes at the cost of write-endurance in practical applications. This apparent long read-after-delay observed for $C_{off}$ is fundamentally different in nature than what is typically observed for Si FeFET, caused by the neutralization of charged shallow traps [18].

The implication of the observation that $C_{off}$ gradually reaches $C_{off,ideal}$ despite long read-after-delay is that achieving down-polarized FE domains ($P_r$<0) across the entire HZO area, which is typically considered as the successful ERS condition for FeFETs, is not a necessary condition to achieve $C_{off,ideal}$ for nvCAPs. This is because the intrinsic channel carrier concentration can be low enough that the channel carriers above "non-polarized" FE domains hardly respond to the AC small signal applied during read. In other words, $C_{IWO}$=$C_{fulldep}$ can be achieved above non-polarized FE domain given that the channel carrier concentration is sufficiently low and AC small signal frequency is sufficiently high. In this sense, retention characteristics of $C_{on}$ and $C_{off}$ at different W doping concentration or intrinsic channel carrier concentration could be an interesting research topic for future work.

$C_{on}$, on the other hand, gradually deviates from $C_{on,ideal}$ as the up-polarized FE domains back-switches to non-polarized state over time due to the depolarization field in HZO layer (i.e., $A_{acc,PGM}$ decreases over time). Gradually decreasing number of up-polarized FE domains results in weaker accumulation which ineffectively screens the depolarization field and accelerates $C_{on}$ decay.

Such a claim on $C_{on}$ is evidenced by the fact that ALD IWO nvCAPs with smaller overlap ($W_{ov}$×$L_{ov}$) or larger device area (W×L) or weaker wake-up condition results in poorer $C_{on}$ retention (Fig. 7(c)-(e)). Fig. 7(c) shows overlap-dependent retention results. Larger overlap areas result in better $C_{on}$ retention attributed to higher PGM efficiency strengthening the accumulation in the non-overlap region. Inspired from this observation, we measure the retention of an ALD IWO nvCAP with a multi-TE structure. In this structure, single TE is split into multiple TEs (fingers) with identical size and spatially distributed across the device area, aiming for electrostatically more effective PGM. Fig. 7(c) shows an improvement of $C_{on}$ retention in the multi-TE structure over single-TE structures, highlighting the role of PGM efficiency in $C_{on}$ retention.

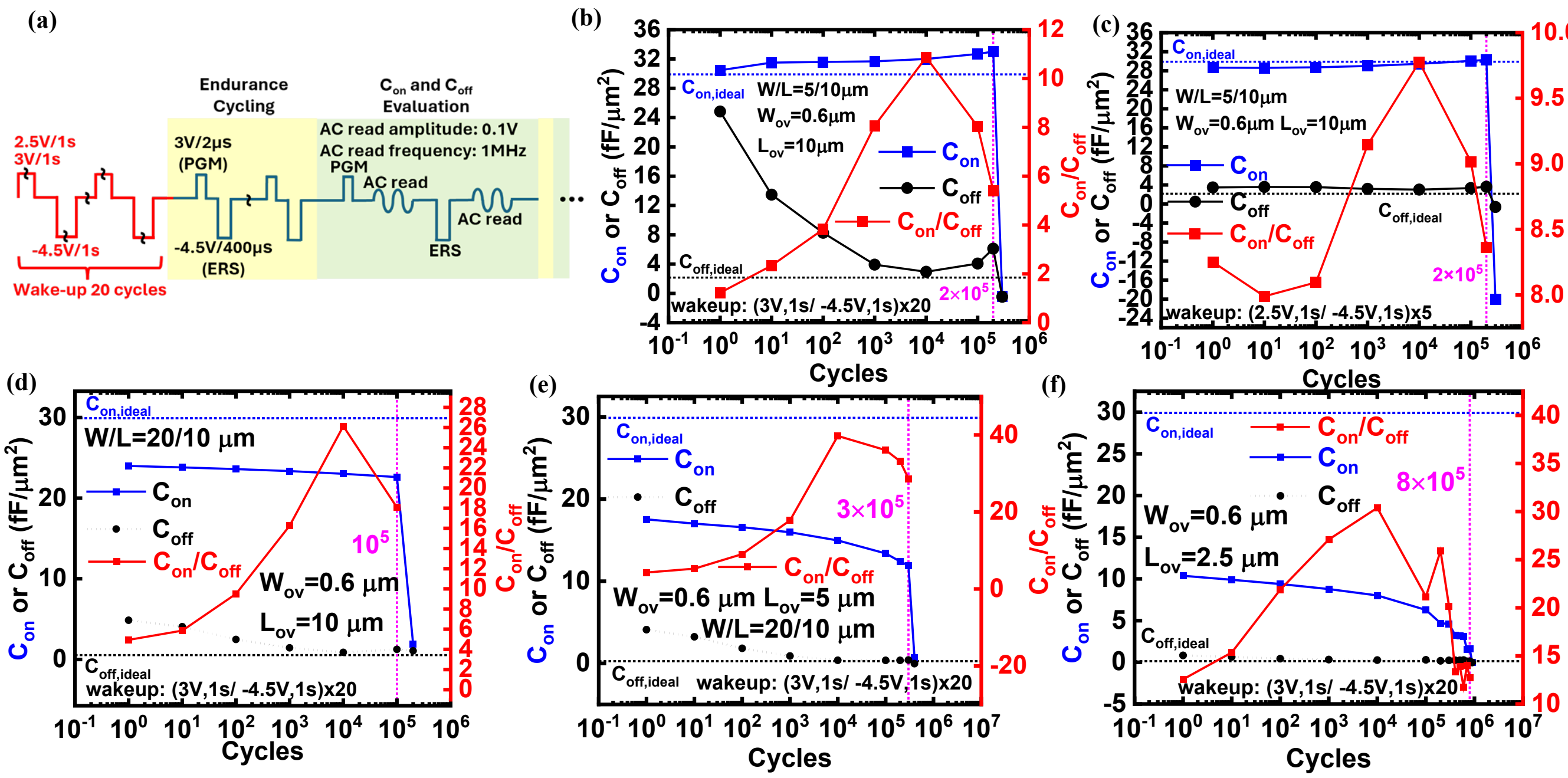

Fig. 8. Measured endurance of an ALD IWO nvCAP at 25 °C under different wake-up conditions with the test setup described in (a). Endurance results obtained with wake-up conditions of (b) (3V,1s/-4.5V,1s)×20 and (c) (2.5V,1s/-4.5V,1s)×5, for a device with W/L=5/10 μm and $W_{ov}/L_{ov}$=0.6/10 μm. Overlap-dependent read endurance results for overlap area conditions of (d) $W_{ov}/L_{ov}$=0.6/10 μm, (e) $W_{ov}/L_{ov}$=0.6/5 μm, and (f) $W_{ov}/L_{ov}$=0.6/2.5 μm, with W/L fixed at 20/10 μm.

Likewise, the device area influences $C_{on}$ retention (Fig. 7(d)). The scaling of device dimension from W=20 to 5 μm naturally leads to more effective PGM, as the TE potential can more effectively penetrate across the channel in the non-overlap region.

Furthermore, Fig. 7(d)-(e) highlights the effect of wake-up conditions on $C_{on}$ and $C_{off}$. Stronger wake-up conditions lead to a larger number of woken up FE domains resulting in more a conductive channel in the non-overlap region, increasing $C_{on}$ retention. At the same time, a larger number of FE domains are partially erased after ERS, becoming the cause of longer read-after-delay.

Lastly, retention characteristics at elevated temperature (85 °C) are examined (Fig. 7(f)). Faster decay of $C_{on}$ and $C_{off}$ compared to 25 °C under an identical device geometry (Fig. 7(e)) can be attributed to FE layer experiencing accelerated back-switching, degraded $P_r$, pronounced imprint at elevated temperature, consistent with prior reports [19-21]. The high gate leakage of the FE layer could also contribute to accelerated retention loss at elevated temperature [22]. Additionally, initial $C_{off}$ value at t=1 s under the same wake-up condition is significantly higher than 25 °C case, causing significant read-after-delay. This read-after-delay is not improved by increasing the ERS pulse width, in contrast to the 25 °C case. Such phenomenon could be attributed to the increased intrinsic carrier concentration of the IWO channel [23]. Higher intrinsic carrier concentration at elevated temperatures makes the ERS more difficult in the non-overlap region increasing $C_{off}$.

Based on the retention analysis of $C_{on}$ and $C_{off}$ under various conditions, several knobs for the stable retention of oxide-channel ferroelectric nvCAPs can be conceived. Retention loss of $C_{on}$ can be minimized by engineering the depolarization field from the FE layer, especially in the non-overlap region. This can be achieved by increasing the PGM efficiency, for example by redesigning the TE geometry (e.g., multi-TE structure) or scaling the TE and BE area toward nanometer scale. Furthermore, retention-aware gate or channel-stack engineering can be considered to improve intrinsic retention of the FE layer. Read-after-delay of $C_{off}$ can be reduced by adopting a new device structure that ensures more effective ERS in the non-overlap region. Alternatively, intrinsic carrier concentration in the non-overlap region can be selectively optimized to reduce read-after-delay of $C_{off}$ at elevated temperature while still ensuring high $C_{on}$ retention.

### *B. Write Endurance*

Next, we examine the write-endurance characteristics of ALD IWO nvCAPs at 25 °C under different geometry and wake-up conditions. The test sequence is described in Fig. 8(a). The write endurance failure point (highlighted in pink font) is defined as the cycling count where abrupt drop (>300 % decrease from previous cycle count) happens in $C_{on}$ or $C_{off}$, which is a possible indication of dielectric breakdown leading to a sharp increase of the number of oxygen vacancies in HZO [24, 25].

When stronger wake-up conditions are used, $C_{off}$ is significantly larger than $C_{off,ideal}$ during initial write cycles due to the read-after-delay issue of $C_{off}$ explained previously in Section V-A. (Figs. 8(b)-(c)). As the write cycling continues, $C_{off}$ tends to decrease due to (i) the increased ERS efficiency with increasing $P_r$ or (ii) gradual back switching of up-polarized FE domains in the non-overlap area after strong wake-up pulses. After $>10^4$ cycles, $C_{off}$ tends to increase possibly due to positively charged defect generation at the interface, until it abruptly drops with $C_{on}$ due to dielectric breakdown.

Figs. 8(d)-(f) show that the write endurance degrades with increasing overlap area. Together with Fig. 7(e), this indicates the trade-off between pulse-mode write-endurance and retention (showing the opposite trend with respect to the overlap area) should be carefully considered for the ALD IWO nvCAP design.

Furthermore, the effect of geometry scaling on write endurance can be understood from the comparison between Figs. 8(b) and (d). Reduction of the device area (=W×L) while maintaining the overlap geometry leads to an enhancement of the write-endurance ($10^5 \rightarrow 2\times10^5$). The overall pulse-mode write, retention and write-endurance results suggest the scaling of device area, combined with overlap area (=$W_{ov}\times L_{ov}$) scaling is expected to simultaneously enhance, $C_{on}/C_{off}$, retention and write-endurance performance of ALD IWO nvCAPs.

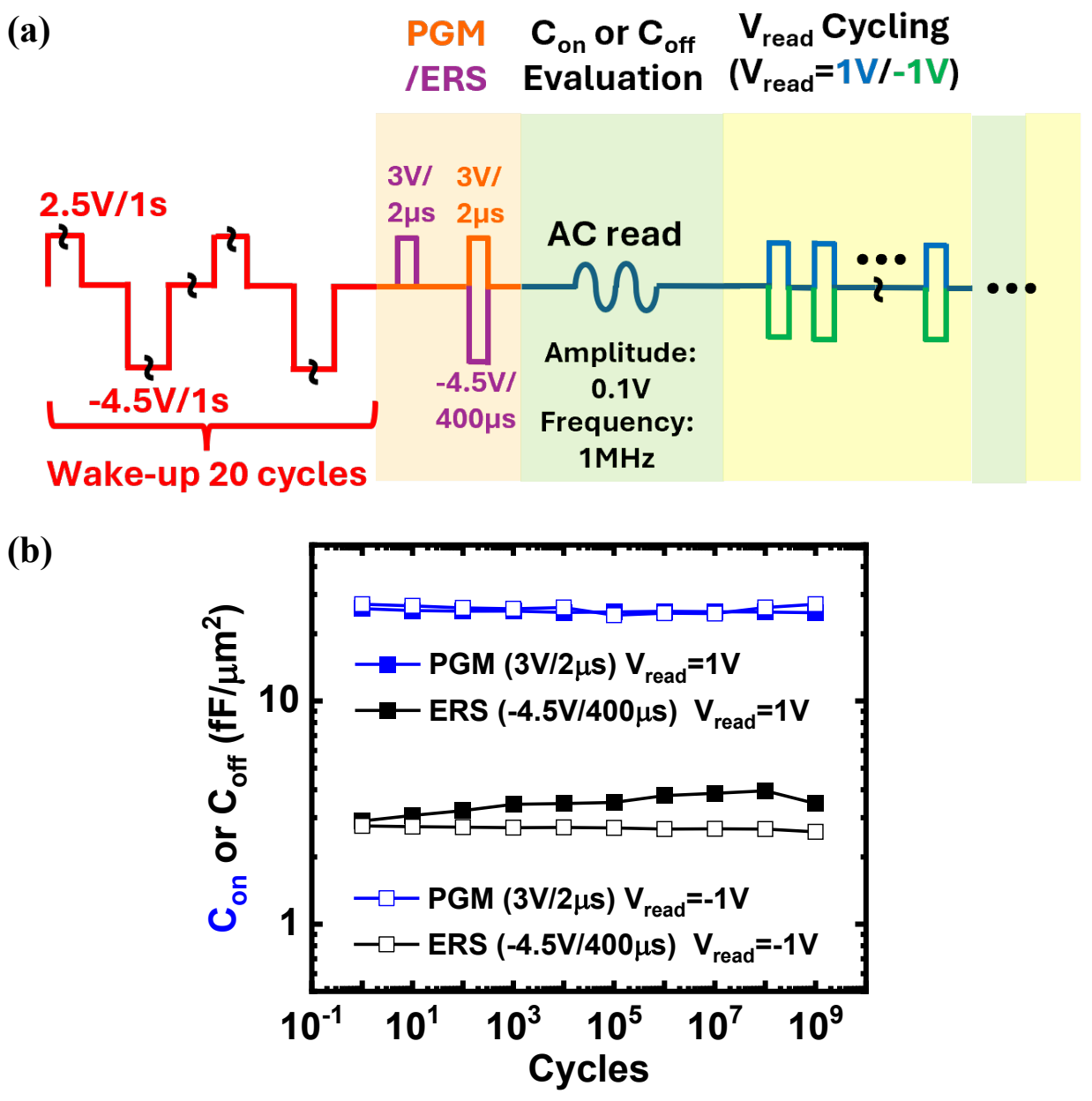

Fig. 9. (a) Read endurance test setup and (b) demonstration of NDRO of $>10^9$ read cycles with $|V_{read}|$=1V at 25 °C. The geometry condition of the ALD IWO is W/L=5/10 μm, $W_{ov}/L_{ov}$=0.6/10 μm.

### *C. Read Endurance*

Lastly, we examine the non-destructive read operation (NDRO) characteristics of ALD IWO nvCAPs through read-endurance measurement outlined in Fig. 9(a). Read stress pulses ($V_{read}$) of ±1V/1μs are applied to "BE" of ALD IWO nvCAP while TE is grounded. Smallest device area (W/L=5/10 μm, $W_{ov}/L_{ov}$=0.6/10 μm) with stable retention until $10^3$ s is chosen for the NDRO test to minimize the impact of retention loss of $C_{on}$ during the read endurance cycling. Considering the read-after-delay of ~100 s for $C_{off}$, the initial $C_{off}$ is measured 100 s after write (ERS) pulse application, while initial $C_{on}$ is measured instantly after write (PGM) pulse application (with <1 s delay). The results shown in Fig. 9(b) indicate that $C_{on}/C_{off}$ does not degrade after $>10^9$ stress cycles. Meanwhile, the voltage polarity dependent read-endurance of $C_{off}$ can be observed. $V_{read}$=-1 V yields better read-endurance than $V_{read}$=1 V for $C_{off}$ while $C_{on}$ shows similar behavior for both polarities. This could be due to the positive built-in field (BE to TE) created by the work function difference between Pd (~5.12 eV) and TiN (~4.5 eV) leading to comparably unstable $C_{off}$ against positive voltages. While the results demonstrate NDRO of $>10^9$ read stress cycles, they simultaneously imply there is a desired TE/BE configuration of ALD IWO nvCAPs; To maximize read endurance, read voltage pulses needs to be applied to "TE" in a practical array implementation with positive operating voltage.

## VII. Conclusion

In this study, we have provided a comprehensive study on the pulse-mode writing and reliability (retention, read endurance, write endurance) characteristics of oxide-channel ferroelectric nvCAPs. We have outlined the governing physics that lead to non-ideal behaviors of oxide-channel ferroelectric nvCAPs, which provides a foundation for understanding their pulse-mode operation and reliability characteristics. Using an ALD IWO nvCAP with asymmetric TE/BE, we have shown that pulse-mode writing characteristics significantly depend on wake-up and write pulse conditions, suggesting the need to optimize both conditions to target desired $C_{on}$ and $C_{off}$ values. Further, pulse-mode retention results have highlighted (i) the role of depolarization field in $C_{on}$ retention and (ii) the impact of inherent hole deficiency of the oxide channel on the pronounced read-after-delay of $C_{off}$. Together with the pulse-mode write endurance results, they have indicated the importance of overlap area optimization and device area scaling to simultaneously improve $C_{on}/C_{off}$, retention and write endurance. Lastly, voltage polarity-dependent NDRO has been examined, demonstrating NDRO for $>10^9$ stress cycles of $|V_{read}|$=1 V while suggesting optimal device configuration for practical array implementation. The study has offered foundational insights into the underlying device physics governing the pulse-mode operation of oxide-channel ferroelectric nvCAPs. These insights are expected to guide the design and operation of oxide-channel ferroelectric nvCAPs toward realizing pragmatic BEOL-compatible non-volatile capacitive arrays.

## Acknowledgment

This work was supported by PRISM and CHIMES, two of the semiconductor research corporation (SRC) centers.